\PassOptionsToPackage{draft}{hypref}
\PassOptionsToPackage{breaklinks=true}{hypref}
\documentclass[a4paper,fleqn,usenatbib]{mnras}
\PassOptionsToPackage{draft}{hyperref}
\usepackage{graphicx}
\usepackage{amsmath}
\usepackage{longtable}
\usepackage{supertabular}

\makeatletter
\@fleqnfalse
\@mathmargin\@centering
\makeatother

\usepackage{color}
\usepackage{natbib}

\newcommand{\RNum}[1]{\uppercase\expandafter{\romannumeral #1\relax}}

\newcommand{\REobs}{\ensuremath{\epsilon}}

\newcommand{\uae}{\ensuremath{\bar{\mu}_{e}}}
\newcommand{\uaeint}{\ensuremath{\bar{\mu}_{e,\mathrm{int}}}}
\newcommand{\uaer}{\ensuremath{\bar{\mu}_{e,r}}}
\newcommand{\re}{\ensuremath{\bar{r}_{e}}}
\newcommand{\rer}{\ensuremath{\bar{r}_{e,r}}}
\newcommand{\reint}{\ensuremath{\bar{r}_{e,\mathrm{int}}}}
\newcommand{\nser}{\ensuremath{n}}
\newcommand{\REpar}{\ensuremath{\REobs(\uaeint,\reint)}}  

\newcommand{\MTObjects}{{\scshape MTObjects}}
\newcommand{\ProFound}{{\scshape ProFound}}
\newcommand{\DeepScan}{{\scshape DeepScan}}
\newcommand{\GALFIT}{{\scshape GALFIT}}
\newcommand{\SExtractor}{{\scshape SExtractor}}
\newcommand{\imfit}{{\scshape imfit}}
\newcommand{\THELI}{{\scshape THELI}}

\newcommand{\remin}{3.0}
\newcommand{\remax}{8.0}
\newcommand{\SBmin}{24.0}
\newcommand{\SBmax}{26.5}

\newcommand{\SBsel}{\SBmin$\leq$\uaer$\leq$\SBmax}
\newcommand{\resel}{\remin\arcsec$\leq$\rer$\leq$\remax\arcsec}

\newcommand{\cutoutsize}{400}

\newcommand{\respread}{6.5}

\newcommand{\reminsynth}{1}
\newcommand{\remaxsynth}{25}
\newcommand{\uaeminsynth}{22}
\newcommand{\uaemaxsynth}{30}
\newcommand{\qminsynth}{0.1}
\newcommand{\qmaxsynth}{1}
\newcommand{\nminsynth}{0.2}
\newcommand{\nmaxsynth}{2.5}

\newcommand{\remto}{{\tt R\_e}}
\newcommand{\magmto}{{\tt mag}}

\newcommand{\kidsarea}{180}
\newcommand{\UnmaskedArea}{39}

\newcommand{\Nsources}{212}
\newcommand{\Nsourcesfull}{829}
\newcommand{\Nfits}{2$\times$10$^{6}$}
\newcommand{\UDGdensity}{8$\pm$3$\times10^{-3}$cMpc$^{-3}$}
\newcommand{\UDGdensitySAM}{2$\times10^{-2}$cMpc$^{-3}$}
\newcommand{\UDGdensityHI}{1.5$\pm$0.6$\times10^{-3}$cMpc$^{-3}$}
\newcommand{\UDGefficiency}{$\sim$0.8$\pm$0.2}

\author[D. J. Prole et al.]{D. J. Prole,$^{1, 2}$\thanks{proled@cardiff.ac.uk}
	R. F. J. van der Burg,$^{2}$
	M. Hilker,$^{2}$
	J. I. Davies$^{1}$
	\\
	$^{1}$School of Physics and Astronomy, Cardiff University, The Parade, Cardiff, CF243AA, UK \\
	$^{2}$European Southern Observatory, Karl-Schwarzschild-Str. 2, 85748 Garching bei M\"unchen, Germany
}

\title[Ultra-Diffuse Galaxies in the Field]{Observational Properties of Ultra-Diffuse Galaxies in Low Density Environments: Field UDGs are Predominantly Blue and Starforming}

\date{}

\pubyear{2018}

\begin{document}
	\label{firstpage}
	\pagerange{\pageref{firstpage}--\pageref{lastpage}}
	\maketitle


\begin{abstract}


While we have learned much about Ultra-Diffuse Galaxies (UDGs) in groups and clusters, relatively little is known about them in less-dense environments. More isolated UDGs are important for our understanding of UDG formation scenarios because they form via secular mechanisms, allowing us to determine the relative importance of environmentally-driven formation in groups and clusters. We have used the public Kilo-Degree Survey (KiDS) together with the Hyper Suprime-Cam Subaru Strategic Program (HSC-SSP) to constrain the abundance and properties of UDGs in the field, targeting sources with low surface brightness (\SBsel) and large apparent sizes (\resel). Accounting for several sources of interlopers in our selection based on canonical scaling relations, and using an empirical UDG model based on measurements from the literature, we show that a scenario in which cluster-like red sequence UDGs occupy a significant number of field galaxies is unlikely, with most field UDGs being significantly bluer and showing signs of localised star formation. An immediate conclusion is that UDGs are much more efficiently quenched in high-density environments. We estimate an upper-limit on the total field abundance of UDGs of \UDGdensity\ within our selection range. We also compare the total field abundance of UDGs to a measurement of the abundance of H\RNum{1}-rich UDGs from the literature, suggesting that they occupy at least one-fifth of the overall UDG population. The mass formation efficiency of UDGs implied by this upper-limit is similar to what is measured in groups and clusters.

\end{abstract}



\section{Introduction}

According to the hierarchical model of galaxy formation \citep{White1978, Kauffmann1993, Cole2000}, dwarf galaxies ($M_{*}\leq\sim10^{9}M_{\odot}$) have constituted the most numerous population of galaxies over cosmic time. Despite this, their small physical sizes and relatively faint brightness can cause them to be proportionally overlooked in observational studies due to difficulties in either detecting them, or distinguishing them from background sources \citep[e.g.][]{Disney1976, Davies2016, Williams2016}. 

\indent Low surface brightness (LSB) galaxies are typically dwarf galaxies in terms of both their stellar and halo mass \citep[e.g.][]{Prole2019}, but have a much lower density of stars. Traditionally, they have been defined by having surface brightnesses at least one magnitude fainter than the night sky, around 22.5 magnitudes per square arc-second in the $g$-band. However, galaxies with much lower surface brightness than this are know to exist \citep[e.g.][]{McConnachie2012, Mihos2015}. LSB galaxies were first proposed to exist \citep{Disney1976} and identified several decades ago \citep{Bothun1987, Impey1988} and have been a subject of scientific discussion ever since \citep[e.g.][]{Davies1989, McGaugh1996, Dalcanton1997, Conselice2003, Sabatini2003, Roberts2007, Lelli2010}. 

\indent Recently, there has been a resurgence of interest in LSB galaxies thanks to the deep imaging of the Dragonfly telephoto array \citep{Abraham2014}, with which \cite{VanDokkum2015} discovered a high abundance of large LSB galaxies in the Coma Cluster. The authors coined the term ``Ultra-Diffuse Galaxy'' or UDG for such objects, a name which has been widely adopted throughout the literature. Specifically, UDGs are defined as objects comparable in effective (half-light) radii to the Milky-Way ($r_{e,r}$$>$1.5 kpc) but are of much lower surface brightness ($\bar{\mu}_{e,r}$$>$24, where $\mu$ denotes a surface brightness in magnitudes per square arc-second and $\bar{\mu}_{e}$ is the average surface brightness within the effective radius). 

\indent There has been much debate over the significance of UDGs and as to whether they make up a different population (in terms of their formation mechanism and thus intrinsic properties) to other, smaller low surface brightness galaxies. There is a growing consensus that UDGs share a continuum of properties with less extreme galaxies in terms of star formation rate \citep{Leisman2017}, size and luminosity \citep{Conselice2018, Danieli2018}, metallicity \citep[e.g.][]{Fensch2018} and perhaps also mass to light ratios \citep[e.g.][]{Prole2019}, however some UDGs may be genuine outliers in the stellar mass - halo mass plane and are devoid of dark matter \citep[e.g.][but see also \cite{Trujillo2018}]{VanDokkum2018, VanDokkum2019}, perhaps suggesting separate formation mechanisms.

\indent It is likely that the popularity of UDGs among the literature is thanks in-part to their large sizes; this makes them easier to identify against background objects in groups and clusters. Indeed, this property has been exploited by several authors in their studies of UDGs in such environments \citep[e.g.][]{Koda2015, Munoz2015, Yagi2016, vanderBurg2016, vanderBurg2017, Janssens2017, Venhola2017, Zaritsky2019, Pina2019}. This bias towards studies in dense environments is exacerbated by the difficulty of obtaining spectroscopic redshifts (and therefore distances) for large samples of LSB galaxies. However, one recent development suggests that it may be possible to estimate distances to such galaxies by exploiting the ubiquity of the globular cluster luminosity function \citep{Roman2019}\footnote{Although it is not currently clear whether the globular cluster luminosity function is ubiquitous for UDGs \citep[e.g.][]{VanDokkum2018, VanDokkum2019}.}. 

\indent UDGs in clusters are typically on the red sequence \citep{Koda2015, vanderBurg2016, Pranjal2019} and show little evidence for tidal interaction even close to the cluster centres \citep{Mowla2017}, suggesting relatively high mass-to-light ratios. There is tentative evidence that UDGs tend to be bluer towards the outskirts of galaxy groups \citep{Roman2017b,  Alabi2018, Zaritsky2019} \citep[and more generally in lower-density environments cf.][]{Greco2018a, Greco2018b}, suggesting that interactions with the environment during in-fall can diminish star formation in the UDGs. While this is not surprising if UDGs are quenched during the in-fall, a separate analysis by \cite{Roman2017a} did not show a significant trend between environmental density and colour.

\indent One outstanding question regarding UDGs is whether they are able to form more efficiently in dense environments like groups and clusters, or whether density plays a detrimental role in UDG formation/survival efficiency. \cite{vanderBurg2017} find that UDGs are relatively more common in higher-mass environments, but this is in tension with other studies \citep{Roman2017b, Pina2018} that came to the opposite conclusion.

\indent Whatever the role of environment in UDG production, several authors \citep[e.g.][]{vanderBurg2016, Whittmann2017, Pina2018} have observed a relative dearth in their number density towards the centres of massive clusters. This suggests that in very high density regions, either UDGs are destroyed quickly or do not form as efficiently. One scenario suggested by \cite{Janssens2017} is that UDGs dissolve in cluster cores, possibly depositing ultra-compact dwarf galaxies in the process.

\indent UDGs can also be understood from a theoretical point of view. UDG formation scenarios can be broadly classified as ``in-situ'' (i.e. secular formation in the absence of interactions with an exterior body or bodies), or environmentally driven. \cite{DiCintio2017} showed through zoom-in cosmological simulations that gas outflows caused by internal feedback processes can produce UDGs within dwarf-sized halos. Further, \cite{Amorisco2016} argue that UDGs can form in-situ both in the field and in cluster environments, and should be expected to do so as the high angular momentum tail of the dwarf galaxy distribution. The importance of high-spin halos for UDG production was also noted by \cite{Rong2017} in their simulations. Indeed, there is much observational evidence suggesting UDGs reside in dwarf sized halos \citep[e.g.][]{Beasley2016b, Amorisco2018, Lim2018, Prole2019}.

\indent However, there are several other feasible formation mechanisms that involve the transformation of normal dwarf galaxies to UDGs though environmental effects. One example is tidal heating, whereby galaxy-galaxy interactions cause an expansion of the dwarfs \citep{Collins2013, Carleton2018}. There are several pieces of observational evidence showing that some UDGs are associated with tidal interactions \citep{Whittmann2017, Bennet2018}. Continually, ram-pressure stripping from the dense intra-cluster medium in groups \& clusters is perhaps able to produce UDGs by quenching early in-fall galaxies \citep{Yozin2015}. \cite{Jiang2018} argue that ram-pressure stripping is the primary effect that causes UDGs to lose gas (and therefore shutdown star formation) in dense environments. This can account for the red colours observed for UDGs in clusters.

\indent While much is known about UDGs in dense environments, relatively little is known about the field population\footnote{We note that our working definition of the field is a representative piece of the Universe in which galaxy groups and clusters are included, but massive haloes naturally make up a relatively small fraction by mass.}, expected to form preferentially through secular mechanisms\footnote{Although for the present study we cannot rule out all external processes such as accretion from gas clouds.} \citep[see, for example, the work of][regarding the evolution of early type dwarf galaxies]{Graham2017, Janz2017}. This is mainly because of the difficulties involved in measuring distances to large samples of LSB galaxies without prior information such as cluster association. Observational studies in groups and clusters alone have been unable to disentangle the relative importance of in-situ vs. environment-driven formation because of the need to perform a statistical background subtraction of interloping (i.e. non-group or cluster) UDG candidates. Some studies \citep{Das2013, Leisman2017, Papastergis2017, Greco2018a, Zaritsky2019} have shown that a field population of UDGs does indeed exist, yet the global properties of these galaxies are poorly understood.

\indent One particularity relevant piece of work is that of \cite{Leisman2017} and \cite{Jones2018}, who have shown that not only do H\RNum{1}-bearing UDGs exist in the field \citep[as theoretically predicted by][]{DiCintio2017}, but also that their number density is too high to be explained by an extrapolation of the empirical relation between the number of UDGs and $M_{halo}$ measured by \cite{vanderBurg2017}. Further, their sample appear systematically bluer than anticipated for UDGs in the field when compared with semi-analytic models \citep{Rong2017, Jones2018}. However, this analysis was limited to H\RNum{1}-rich field UDGs and it is unknown how this population relates to the overall field UDG population.

\indent In this work, we use deep, wide-field imaging combined with an empirical UDG model to statistically constrain the global properties of UDGs in the field without knowing the distances to any of our sources. This includes an analysis of their colours, number density and mass-formation efficiency. The paper is structured as follows: We describe our data in $\S$\ref{section:data}. We describe our sample of UDG candidates in $\S$\ref{section:measurements} and quantify our recovery efficiency. In $\S$\ref{section:predict} we describe our empirical model of UDGs and potential interlopers in our UDG candidate sample. Our results are presented and discussed in sections \ref{section:results} and \ref{section:discussion} respectively. We conclude in $\S$\ref{section:conclusion}. All magnitudes are quoted in the AB magnitude system. Cosmological calculations are performed assuming $\Lambda$CDM cosmology with $\Omega_{\mathrm{m}}$=0.3, $\Omega_{\Lambda}$=0.7, H$_{0}$=70 kms$^{-1}$Mpc$^{-1}$.



\section{Data}
\label{section:data}

For source detection and structural parameter estimation, we use a \kidsarea\ deg$^{2}$ subset of data from the Astrowise \citep{McFarland2011} reduction of the Kilo-Degree Survey \citep[KiDS;][]{deJong2013, Kuijken2019} that overlaps with the GAMA spectroscopic survey \citep{Driver2011} equatorial fields. We use the $r$-band for source detection because it is the deepest and has the best image quality. This is the same data\footnote{Although they use the \THELI\ \citep{Erben2013} KiDS reduction, the depth is essentially equivalent.} used by \cite{vanderBurg2017} in their study of the UDG populations of galaxy groups and so we can make direct comparisons with their findings. Despite the GAMA overlap, redshift measurements are not available for most of our sources because they are generally much fainter than the limiting depth of GAMA at 19.8m$_{r}$.

\indent The pixel size of KiDS is 0.2$\arcsec$, small enough to properly sample the point spread function (PSF) that has a typical full-width at half-maximum FWHM$<$1$\arcsec$. The sky-background is estimated in meshes of 20$\arcsec$ that are median filtered over in 3$\times$3 meshes.

\indent While the KiDS $r$-band is sufficient to reach a limiting surface brightness of \uaer$\sim$26.5, we additionally use the first data release of the overlapping Hyper-Suprime-Cam Subaru Strategic Program \citep[][]{Aihara2018} to measure colours. The HSC-SSP data is around 0.5m$_{r}$ deeper than KiDS (more so in the $g$-band), but has a smaller overlapping footprint by about a quarter compared to the KiDS area we consider. This leaves us with $\sim$\UnmaskedArea\ deg$^{2}$ of unmasked data from which we can measure HSC-SSP colours. 

\indent Compared to the  \kidsarea\ deg$^{2}$ KiDS-GAMA overlap we use here, the remaining footprint that we have HSC-SSP data for is fairly limited. This may make us partially sensitive to cosmic variance. However, we note that our footprint is spread uniformly over three GAMA regions (G09, G12 and G15), each separated by at least 26 degrees. Additionally, we can account for local galaxy groups and clusters using the GAMA group catalogue \citep[][see $\S$\ref{section:groups}]{Robotham2011}. In the future, our analysis can be easily upscaled to larger footprints.

\indent We note that we do not use the HSC-SSP for detection because of its limited footprint and because its background subtraction is more aggressive compared to KiDS (mesh grid of $\sim$20$\arcsec$ but with no median filtering over meshes), meaning that it could restrict the maximum angular size of sources we could measure accurately. For the present analysis, we restrict ourselves to the $g$ and $r$ bands but note that this can be expanded in future studies.


\begin{figure*}
	\includegraphics[width=\linewidth]{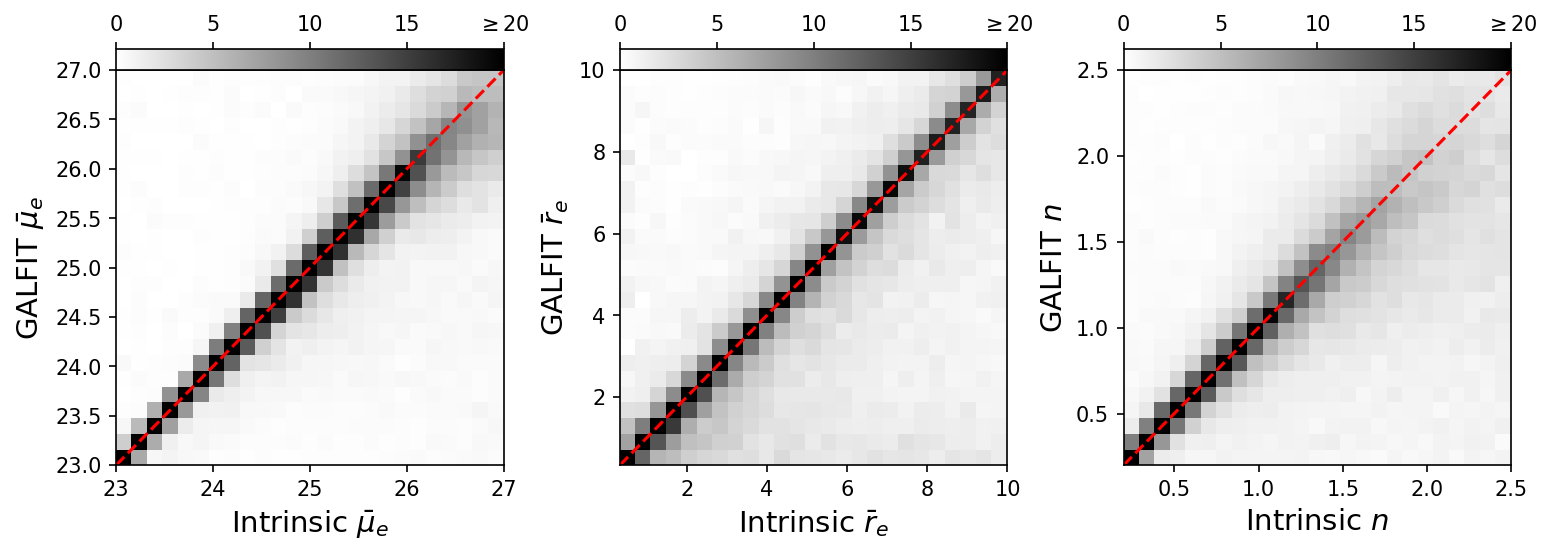}
	\centering
	\caption{This figure shows how the measured \GALFIT\ parameters compare with the intrinsic S\'ersic parameters of the synthetic galaxies that we inject into the data. The black 2D-histograms show this data for our sample of recovered injections. The red dashed line is the one-to-one relation and is not a fit. Clearly we are able to recover the intrinsic parameters with good accuracy and precision over the range of parameter space that we are interested in. However, our precision diminishes slightly when recovering high $n$ (S\'ersic index) sources, but we retain accuracy such that there is little bias in the recovered parameter estimates. The units of the histogram are the percentage of sources with intrinsic parameters (given by the column) that occupy a particular bin in measured parameter space. The colour map is capped at 20\% in order to increase the contrast for bins of lower completeness.}
	\label{figure:galfit}
\end{figure*}


\section{Measurements}
\label{section:measurements}

Since we do not know the distance to any of our sources (apart from a small subset, cf. appendix \ref{appendix:redshift}), we must rely on selection criteria defined in observable parameter space (i.e. that which we measure as projected on the 2D surface of the sky). Specifically, we target the LSB regime $\SBmin\leq\uaer<\SBmax$, where the lower limit is chosen for consistency with the literature and the upper limit is defined by the depth of KiDS. The upper selection limit on \re\ is chosen to be much smaller than the spatial scale of the KiDS background subtraction and we set it as \remax\arcsec\ in line with \cite{vanderBurg2017}. The lower limit on \re\ is more difficult to set; while technically we are limited by the size of the PSF (FWHM $\sim$0.6\arcsec), it is also worth considering that the number of contaminant sources (i.e. non-UDGs) that satisfy our selection criteria quickly increases as this limit is lowered because of massive galaxies in the background. This issue is compounded by the fact that we do not have the advantage of a directly-measurable background surface density compared to similar studies in groups and clusters. Here, we use a lower limit of \re$\geq$3$\arcsec$ for our selection \citep[e.g.][]{Sabatini2003, Davies2016}. At a redshift of $z$=0.2,  \re=3$\arcsec$ corresponds to $\sim$10 kpc. Our upper limit of \re=8$\arcsec$ corresponds to 1.6 kpc at $z$=0.01.
 

\subsection{Source detection and measurement}
\label{section:detection}

\indent We choose to improve upon the catalogue used in \cite{vanderBurg2017}, who used \SExtractor\ for source extraction, by using software optimised for the detection of LSB sources; this enables us to probe slightly deeper than their catalogue. We have experimented with several different detection and segmentation algorithms, including \MTObjects\ \citep{Teeninga2016}, \ProFound\ \citep{Robotham2018}\footnote{https://github.com/asgr/ProFound} and \DeepScan\ \citep{Prole2018}\footnote{https://github.com/danjampro/DeepScan}. After some consideration, we selected \MTObjects\ as the most suitable for our analysis because it seemed to produce less spurious detections around large, bright galaxies in our pipeline. We note that during this work \ProFound\ has been updated with an alternative segmentation algorithm that improves its reliability around such objects, but we have not tested this. We used default parameters from  \MTObjects: $\alpha$=$10^{-6}$ and {\tt move\_factor}=0.5, where $\alpha$ sets the statistical significance level for the de-blending, and {\tt move\_factor} determines the spread of large objects.

\indent We used the KiDS weight images to mask out regions in the data which have less than three exposures contributing to the imaging prior to the \MTObjects\ run. This was done to ensure uniform sensitivity over the full data set, as \MTObjects\ relies on a global estimate of the background distribution.


\subsubsection{Point spread function measurement}

\indent We took advantage of our decision to split the KiDS frames into 3$\times$3 subframes by making one PSF model per subframe (i.e. nine PSF models per square degree). This was accomplished by targetting point sources in the  \remto\ and \magmto\ plane (\MTObjects\ estimates of the effective radius and total magnitude respectively) based on our \MTObjects\ catalogues from each subframe. Point sources were required to have an axis ratio as estimated by \MTObjects\ greater than 0.9. We then fit Moffat profiles to the individual point source candidates using \GALFIT\ \citep{Peng2002}. Our final PSF model for each subframe was taken as the model corresponding to the mean Moffat FWHM after a sigma-clipping algorithm was applied to remove outliers. We measure a mean FWHM of 0.6$\arcsec$ with a standard deviation of 0.1$\arcsec$ over the full KiDS area that we use. We measure a median value of the Moffat $\beta$ parameter of 2.2, with a standard deviation of 0.1.

\subsubsection{Source extraction pipeline}

Following measurement of the PSF, our overall detection and measurement consists of the following steps:

\begin{enumerate}
\item Use \MTObjects\ to produce a segmentation image and preliminary source catalogue for each subframe.

\item Apply a pre-selection to the preliminary catalogue to identify candidates suitable for input to \GALFIT. This is necessary to ensure a practically feasible number of fits. Specifically, we required \texttt{R90}$>$1.5$\arcsec$ and  \texttt{mu\_mean}$>$23.5 mag arcsec$^{-2}$, where \texttt{R90} and \texttt{mu\_mean} are a proxy for the radius containing 90\% of the galaxy light and the average surface brightness, respectively.

\item Use \GALFIT\ to fit a combined S\'ersic plus inclined sky plane model to each pre-selected source, ignoring masked pixels and additionally masking other segments from the \MTObjects\ segmentation images. Parameter estimates from \MTObjects\ were used as the initial guesses for \GALFIT. The sky RMS is estimated directly from pixels in the cut-out region that were unmarked in the segmentation image. The S\'ersic profile is defined as,

\begin{equation}
I(r)= I_{e}\exp\left\lbrace -b_{n} \left[ \left(\frac{r}{r_{e}}\right)^{\frac{1}{n}}-1\right] \right\rbrace
\end{equation}

\noindent where $I(r)$ is the galaxy's intensity as a function of radius, $I_{e}$ is the intensity at the effective (half-light) radius $r_{e}$, and $b_{n}$ is a constant determined only by the index $n$, which in turn governs the profile slope. We note that all conversions between S\'ersic parameters are performed using the prescriptions of \cite{Graham2005}.

\item Apply the selection criteria to the resulting \GALFIT\ models in order to produce a final catalogue of UDG candidates.
\end{enumerate}

\noindent The PSF models were used as an input to \GALFIT\ such that the measurements correspond to deconvolved parameters. The \GALFIT\ cut-out size was \cutoutsize\ pixels, large enough to recover the intrinsic S\'ersic parameters properly over the full range of parameter space we explore here, and was tested with the synthetic source injections (see $\S$\ref{section:RE}). We note that it was important to include the inclined sky plane component in the \GALFIT\ modelling in order to retrieve unbiased measurements in the cases of high \re\ or high \nser\ profiles.

\indent Following the full set of pre-selections, we are left with $\sim$\Nfits\ sources that are input to \GALFIT\ over the full \kidsarea\ degree squared KiDS area. Following the selection using  \GALFIT\ parameters (\SBsel, \resel, $n$$\leq$2.5), we are left with \Nsourcesfull\ UDG candidates. After selecting sources also within the HSC-SSP footprint, our final catalogue of UDG candidates consists of \Nsources\ sources; some examples are shown in appendix \ref{appendix:examples}. We note that contrary to UDGs in clusters and groups, our sample comprises of sources that appear far more irregular, with features suggestive of active star formation.


\subsection{Recovery Efficiency}
\label{section:RE}

\indent We define the recovery efficiency \REobs\ as the fraction of sources with intrinsic observable parameters (i.e. without the effects of measurement uncertainty) that have measurements that meet our selection criteria. As such, sources that do not meet our selection criteria in terms of their intrinsic observable parameters may be selected (\REobs$>$0) because of measurement uncertainty. Anti-correlated with \REobs\ is therefore the selection purity, defined as the fraction of detections with intrinsic observable parameters that do not meet the selection criteria, but have measured properties that do and thus make it into the final catalogue of UDG candidates. This is different compared to the purity of the UDG candidates, which is defined as the fraction of sources in the UDG candidate catalogue that are intrinsically UDGs as defined by their physical properties.

\indent We have used synthetic source injections to quantify \REobs\ as a function of intrinsic S\'ersic parameters, \REpar. To do this, we create mock images by inserting artificial galaxy profiles (PSF-convolved, one-component S\'ersic) into each frame of the real KiDS data and run them through the full detection and measurement pipeline described in $\S$\ref{section:detection}. The S\'ersic parameters were drawn uniformly from the ranges presented in table \ref{table:RE}, where $q$ is the observed axis ratio. As noted by \cite{vanderBurg2016}, \re\ (the circularised half-light radius) is robust against the intrinsic distribution of axis ratios and \uae\ is a better indicator of a sources detectability than other parameters such as the central surface brightness. Note that we do not include the S\'ersic index $n$ as a free parameter in \REpar\ in order to simplify the analysis. This does not severely impact our results because the intrinsic range in $n$ for UDGs is narrow (e.g. \cite{Koda2015} find a mean of $n$$\sim$1 with a standard deviation of 0.34).

\indent The sources were injected at a surface density low enough to ensure that  injected profiles were separated on average by \respread\ times the maximum value of \re\ given in table \ref{table:RE}. We repeated the process several times in order to increase the number statistics for the \REpar\ measurement, simulating $\sim$735,000 sources overall.

\begin{table}
\begin{tabular}{ c c c }
\hline
  Parameter & Lower limit & Upper limit \\
\hline
  \uae\ [mag arcsec$^{2}$] & \uaeminsynth & \uaemaxsynth \\
  \re\ [arcsec] & \reminsynth & \remaxsynth \\
  $q$ & \qminsynth & \qmaxsynth \\
  $n$ & \nminsynth & \nmaxsynth \\
\hline
\end{tabular}
\caption{Parameter ranges for the artificial galaxy injections. The parameter realisations are drawn uniformly within the ranges, which are much wider than our selection criteria for UDG candidates.}
\label{table:RE}
\end{table}

\begin{figure}
	\includegraphics[width=\linewidth]{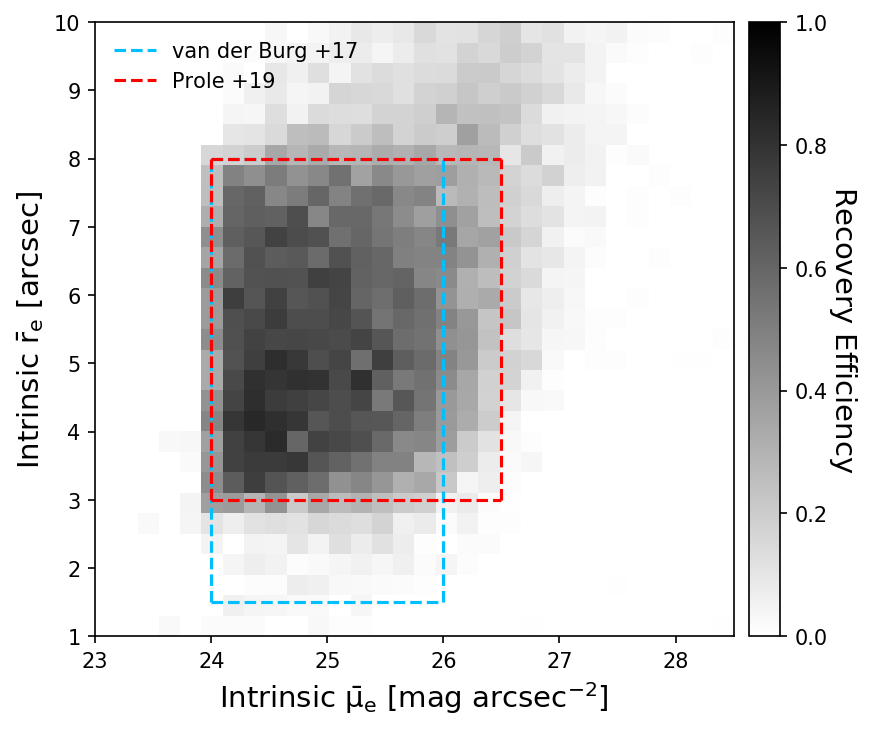}
\centering
\caption{The recovery efficiency of synthetic sources injected into the data as a function of circularised effective radius and mean surface brightness within the effective radius given our selection criteria. The red box indicates our selection criteria, while the blue box is that used in \protect\cite{vanderBurg2017}.}
\label{figure:RE}
\end{figure}

We only considered unmasked sources for the estimate of \REpar, which was measured with $\sim$550,000 artificial galaxy injections spread evenly over our full KiDS subset. We show our ability to precisely measure the intrinsic parameters of our injected sources in figure \ref{figure:galfit}. Our fiducial measurement of \REpar is shown in figure \ref{figure:RE}. 

We note that we additionally tested \imfit\ \citep{Erwin2015} in place of \GALFIT\ in the pipeline described in $\S$\ref{section:detection} and found that it made no significant difference to our measurement of \REpar.

\indent A criticism of the above method is that it relies on field galaxies being well fit by S\'ersic profiles. While this is certainly justified in dense environments where UDGs show little evidence of tidal features \citep{Mowla2017}, it is not clear whether this is justified for the field population. Since field galaxies are expected to be relatively isolated, they are likely to show little evidence of tidal disruption. This does not however rule out irregular morphologies caused by bright star-forming regions and  secular processes like stellar feedback from e.g. supernovae.

\indent In addition to the injections described above and in equal measure, we inject nucleated profiles (S\'ersic + Moffat PSF model) to the data assuming that approximately 1\% of the galaxy light is contained within the nucleus. This allows us to quantify any systematic differences in our recovery efficiency that might be caused by the presence of nuclei and adapt our selection criteria accordingly. We note that we do not attempt to fit nucleated profiles for our measurements because experiments with artificial galaxy profiles showed that the fits were not reliable. We find that the presence of a nucleus is sufficient to positively bias recovered values of the S\'ersic index by approximately 30\% at $n$=1. Since almost all recorded UDGs have S\'ersic indices less than around 1.5 \citep[there is both observational and theoretical evidence for this, e.g.][]{Koda2015, Roman2017b, DiCintio2017,  Venhola2017}, our upper selection limit of $n$=2.5 is justified. The effect of the selection in $n$ on our results is discussed further in $\S$\ref{section:sersicn}.

\indent In comparison to \cite{vanderBurg2017}, our selection criteria allow us to probe $\sim$0.5 magnitudes deeper in surface brightness. The increased depth comes in-part from the proficiency of \MTObjects\ over \SExtractor\ for detecting LSB sources. We also use a higher cut in \re; since we do not have the advantage of a measurable background level, imposing a higher minimum cut in \re\ allows us to mitigate against an excessively contaminated sample of UDG candidates. In figure \ref{figure:RE}, it can be seen that we expect some contamination from (apparently) large, faint sources that do not intrinsically meet our selection criteria (top right of the figure). However, since the number of apparently large sources is very small compared to the number of smaller ones, the decrease in purity from such sources is negligible.

\subsection{Colours}
\label{section:colours_obs}

We exploit the overlap of the KiDS survey with the HSC-SSP footprint in order to measure $(g-r)$ colours for our UDG candidates. We remind the reader that while HSC-SPP is $\sim$0.5 magnitudes deeper than KiDS, the background subtraction in HSC-SPP is slightly more aggressive.

\indent We use an aperture-based strategy to measure $(g-r)$ colours. Specifically, all colours are measured within the 1\re$_{,r}$ ellipses from our \GALFIT\ measurements. We estimate the sky level along with its uncertainty using a random aperture approach, whereby we place 100 equally shaped apertures in close vicinity to (but not touching) the source. Before measuring the median background level and its uncertainty, fluxes are sigma-clipped at 2$\sigma$\footnote{The bias in the recovered standard deviation when sigma-clipping normally distributed data at $2\sigma$ is approximately 25\%; we therefore correct our estimates by this factor.} in order to lower the potential for over-estimating the background level because of nearby sources. We do not perform additional aperture corrections because our sources are much more extended than the PSF.

\indent Due to the increased depth offered by HSC-SSP, we are able to measure positive fluxes in the $g$ and $r$-bands for close to 100\% of our sources. The typical measurement error in $(g-r)$ due to the background fluctuations is approximately 0.04 magnitudes; this is discussed further in $\S$\ref{section:predict_error}. We show comparisons between KiDS and HSC-SSP imaging in figure \ref{figure:examples}.


\section{The empirical model}
\label{section:predict}

Without knowing the distances to any of our sources, it is difficult to tell how many are intrinsically UDGs and how many are cosmologically dimmed background galaxies. In this section, we describe an empirical model that can be used to generate a synthetic population of UDGs in order to compare with observation. This is supported by an additional model for massive galaxies that allows us to estimate the number of non-UDG contaminants in our observational sample.


\subsection{Empirical UDG model}
\label{section:predict_UDG}

\subsubsection{Empirical properties of UDGs}
\label{section:predict_UDG_int}

\indent One of the simplest models can be created by assuming field UDGs share similar empirical properties with UDGs in clusters. Of importance for our analysis are prescriptions for \uae$_{,\mathrm{int}}$ (the surface brightness corrected for cosmological projection effects), \re$_{,\mathrm{phys}}$ (physical size) and $(g-r)_{\mathrm{int}}$, the rest-frame colour. As discussed in $\S$\ref{section:RE}, we assume that all UDGs occupy the range of 0.2 to 2.5 in S\'ersic index.

\indent  \cite{vanderBurg2016} recorded that the distribution of average surface brightness \uae\ is approximately uniform in group environments for UDGs. This has been complimented by the findings of \cite{Danieli2018}, who found that the distribution of absolute magnitude at fixed size is approximately uniform for large, red galaxies in the Coma Cluster after accounting for the newly discovered UDGs. These two observations are equivalent, since at a given size the mean surface brightness is uniquely defined by its magnitude (i.e. there is no dependence on S\'ersic index). However, the work of \cite{Danieli2018} showed that this relation extends from the low surface brightness regime and to much brighter galaxies. We therefore adopt a uniform distribution $U^{x_{\mathrm{max}}}_{x_{\mathrm{min}}}(x)$ for \uae$_{,\mathrm{int}}$:

\begin{equation}
 \bar{\mu}_{e,\mathrm{int}}\sim U^{26.5}_{24.0}(\bar{\mu}_{e,\mathrm{int}})
 \label{equation:sbmodel}
\end{equation}

\indent The subsequent observational study of \cite{vanderBurg2017} \citep[supported theoretically by][]{Carleton2018} has shown that the size distribution of UDGs in groups and clusters is well described by a power law of slope $-2.71 \pm0.33$ in logarithmic size bins, such that smaller UDGs are much more common than larger ones. The intrinsic distribution of physical sizes in kpc is therefore taken as:

\begin{equation}
\bar{r}_{e,\mathrm{phys}}[\mathrm{dex}]\sim \bar{r}_{e,\mathrm{phys}}^{-2.71}
\label{equation:remodel}
\end{equation}

\noindent where we assume the range of $\bar{r}_{e,\mathrm{phys}}$ lies between 1.5 and 7.0 kpc, consistent with \cite{vanderBurg2017}. We probe the effect of varying the power-law slope on our result in appendix \ref{appendix:uncertainty}.

\indent It has been noticed by several authors that UDGs in clusters tend to lie on the red-sequence \citep[e.g.][]{Koda2015, vanderBurg2016} and this is also expected theoretically. There have been hints that UDGs may tend to be much bluer in less-dense environments \citep{Roman2017b, Jones2018, Jiang2018}, although this is not always clear from an observational perspective \citep{Roman2017a}. As such, colours of field UDGs remain relatively poorly understood. We therefore leave the distribution of  $(g-r)_{\mathrm{int}}$ as a variable of our model, and discuss it further in $\S$\ref{section:predict_UDG_colour}.

\begin{figure}
	\includegraphics[width=\linewidth]{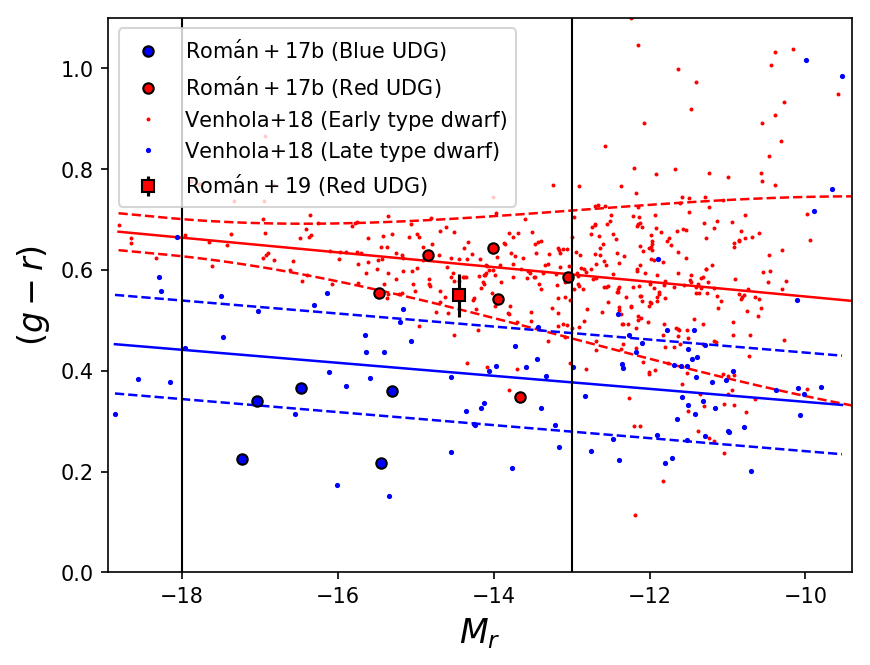}
	\centering
	\caption{Our model colour distributions for the mock UDG catalogues. The data we fit are from \protect\cite{Venhola2018} and correspond to late-type (small blue points) and early-type (small red points) dwarf galaxies in the Fornax cluster. We fit simple linear models (uninterrupted coloured lines) and their 1$\sigma$ uncertainties (dashed lines) after clipping outliers at 3$\sigma$. Under the assumption that UDGs have similar colours to dwarf galaxies, we use the late/early type fits for our star-forming/quiescent mock UDG catalogues. The vertical black lines span the approximate range of absolute magnitudes occupied by UDGs \protect\citep{vanderBurg2016}. The bold points show the red/blue UDGs of \protect\cite{Roman2017b} and \protect\cite{Roman2019}.}
	\label{figure:model_colours}
\end{figure}

\subsubsection{Estimated number density of UDGs}
\label{section:predict_UDG_dense}

\indent We use the (almost linear) empirical relation between the number of UDGs and the mass of their parent halo measured by \cite{vanderBurg2017} to estimate the formation efficiency of UDGs per unit mass in clusters and groups. From this, we can calculate the total number of UDGs that should exist out to redshift  $z_{\mathrm{max}}$, using our cosmological model to estimate the total enclosed mass. We calculate the total mass $M_{\mathrm{tot}}$ contained within the volume $V$ probed by solid angle $\omega$ out to $z_{\mathrm{max}}$ using equation \ref{equation:Mtot},

\begin{equation}
\label{equation:Mtot}
M_{\mathrm{tot}} = \int_{0}^{z_{\mathrm{max}}}\Omega_{m}(z)\rho_{\mathrm{crit}}(z)\frac{dV}{dz\,d\omega}dz\,d\omega
\end{equation}

\noindent where $\Omega_{m}$ is the fractional contribution of matter to $\rho_{\mathrm{crit}}$, the critical density of the Universe. We additionally assume that the UDGs are spatially distributed smoothly according to the integrand of equation \ref{equation:Mtot}, such that the redshift distribution of field UDGs follows the mass. 

\indent Note that the adopted value of $z_{\mathrm{max}}$ does not impact the result, provided that the number of sources we predict to observe out to $z$ (given our recovery efficiency) has converged, i.e. has stopped increasing, before $z_{\mathrm{max}}$. For our modelling we use $z_{\mathrm{max}}$=1, which meets this criterion (see $\S$\ref{section:results_comp}).

If we assume that UDGs form with an average efficiency equivalent to a $10^{15}$M$_{\odot}$ cluster according to equation 1 of \cite{vanderBurg2017}, we derive a volume density of $\sim9\times10^{-3}$cMpc$^{-3}$. This is a factor of six higher than the total number density of H\RNum{1}-bearing UDGs measured by \cite{Jones2018}, who measured a value of \UDGdensityHI. Using a different value than $10^{15}$M$_{\odot}$ for the halo mass would not strongly modify the initial number density estimate since the slope of the relation between $M_{halo}$ and the number of UDGs hosted by the halo is approximately one \citep[at least down to $M_{halo}$$\sim$10$^{12}\mathrm{M_{\odot}}$,][]{Roman2017b, vanderBurg2017, Pina2018}. However, by selecting a halo mass of $10^{15}$M$_{\odot}$, we are essentially comparing the field abundance with that in a  $10^{15}$M$_{\odot}$ cluster in our later analysis. We estimate the impact that the uncertainty in the \cite{vanderBurg2017} relation has on our result in appendix \ref{appendix:uncertainty}. 


\subsubsection{Accounting for cosmological effects}
\label{section:predict_UDG_cosmo}

We account for the cosmological distance modulus, angular diameter distance $d_{a}$ (describing how physical sizes map to angular sizes as a function of the redshift, $z$) and $k$-corrections \citep[the filter and spectral energy distribution (SED) dependent effect that modifies a source's apparent brightness with $z$, independently from the distance modulus,][]{Hogg2002}. In combination, these quantities allow us to project the surface brightnesses and angular sizes of our mock sources out to a certain redshift.

\indent While $d_{a}$ is simple to account for, the exact $k$-correction depends on the assumed SED for the UDGs. Quiescent UDGs are thought to be old, metal poor galaxies \citep[e.g.][]{Ruiz-Lara2018, Ferre-Mateu2018, Fensch2018}. We adopt the average UDG properties from \cite{Ferre-Mateu2018} to estimate the $k$-corrections for such galaxies, namely an age of 6.7 Gyr and [Z/H]=-0.66. In the case of star-forming UDGs, we assume the same age and metallically as in the quiescent model, but introduce star-formation at a uniform rate until the time of observation. While this is an idealised scenario, we probe the significance on the assumed model for $k$-corrections in appendix \ref{appendix:uncertainty}. All stellar population models and $k$-correction estimates are calculated using the Flexible Stellar Population Synthesis \citep[FSPS,][]{Conroy2009, Conroy2010} code. For the KiDS $r$-band, we assume SDSS-like filters for the $k$-correction estimates. For the HSC-SSP colours, we use Subaru Suprime Cam filters.

\indent We construct mock catalogues by sampling intrinsic parameters from the appropriate distributions. Following this, we convert the units into apparent, observed quantities through the cosmological distance modulus, angular diameter distance (for the angular sizes) and band-specific $k$-corrections.

\subsubsection{UDG colour models}
\label{section:predict_UDG_colour}

The inclusion of colour into our analysis is critical because the colour of a galaxy contains some information about its distance thanks to the cosmological redshifting of spectroscopic features. While we have assumed a stellar population model for the UDGs in order to estimate the $k$-corrections, we cannot use these models to assign colours to our mock catalogue because of the need to include some intrinsic scatter. One alternative approach is to model the colours using measurements from the literature. 

\indent Several authors have shown that UDGs occupy the red sequence in clusters \citep[e.g.][]{Koda2015, vanderBurg2016} and this is also supported theoretically \citep{Rong2017}. However, modelling the colour distribution of star-forming UDGs is slightly harder because there is not as much available data for them. Since UDGs have stellar populations similar to dwarf galaxies \citep[e.g.][]{Fensch2018}, one viable method is to assume that star-forming UDGs have colours similar to late-type dwarf galaxies. 

\indent \cite{Venhola2018} have measured the $(g-r)$ colours as a function of absolute magnitude for such galaxies in the Fornax cluster. Using these measurements, it is possible to fit the relationship between colour and absolute magnitude with a simple linear model separately to each of their early and late type samples. For the late type galaxies, we use a constant scatter term, while we interpolate the standard deviation of the colours in bins of absolute magnitude for the early type galaxies. We show the corresponding fits in figure \ref{figure:model_colours}, where we have clipped outliers at 3$\sigma$. We note that our fit to the early-type dwarf galaxies is consistent with the approximate fit to the red-sequence UDGs in clusters from \cite{vanderBurg2016}.

\indent In figure \ref{figure:model_colours} we also compare with the UDGs discovered by \cite{Roman2017b}\footnote{where the $(g-r)$ colours have been kindly provided by Javier Rom\'an.}, which have been decomposed into red and blue populations based on their $(g-i)$ colour. While their red population is fairly consistent with our red-sequence model, the blue UDGs seem to be systematically bluer than our colour model for blue galaxies. This may be explained by the fact that our model is based on measurements from the Fornax galaxy cluster where environmental processes, for example ram-pressure stripping, may cause reddening of the galaxies. In comparison, the UDGs of \cite{Roman2017b} are found in isolated galaxy groups where such effects are less prolific.


\subsection{Empirical model for massive galaxies}
\label{section:predict_interlopers}

Not all of the UDG candidates in our observational sample are intrinsically UDGs. As large, bright galaxies are shifted towards higher redshift, they become both fainter and smaller in terms of their angular size and may eventually satisfy our selection criteria. Equally, small foreground dwarf galaxies not meeting the UDG criteria have the potential to contaminate the sample. Since we are not able to directly measure the number of these interlopers (as can easily be done when considering a group or cluster environment), we are forced to use empirical relations from the literature to estimate the level of contamination.

\indent It is standard practice to broadly categorise galaxies as either late-type or early-type based on their morphology and/or colour \citep[e.g.][]{Bell2003, Baldry2004, Driver2006, Taylor2015}. Massive early-type galaxies (ETGs) are typically quiescent and therefore redder than late-types. Additionally, massive ETGs generally have higher S\'ersic indices compared to late types. For ETGs, the S\'ersic index increases with total stellar mass \citep[e.g.][]{Caon1993, Graham1996, Danieli2018}. While early-type dwarf galaxies exist with  S\'ersic indices around one \citep[e.g.][]{Prole2018}, we are probing the field population and therefore expect that the main contribution from ETG interlopers will be from higher mass galaxies with correspondingly higher S\'ersic indices. We discriminate against recovering such objects in our UDG candidate sample through the upper-limit cut in S\'ersic index at $n$=2.5. As such, we expect the dominant source of contamination in terms of massive galaxies ($M_{*}>10^{9}M_{\odot}$) to be mainly constituted of massive late-type galaxies. By contrast to massive ETGs, late-type galaxies are systematically bluer, with S\'ersic indices $n<$2.5 \citep{Vulcani2014}.

\begin{figure*}
	\includegraphics[width=\linewidth]{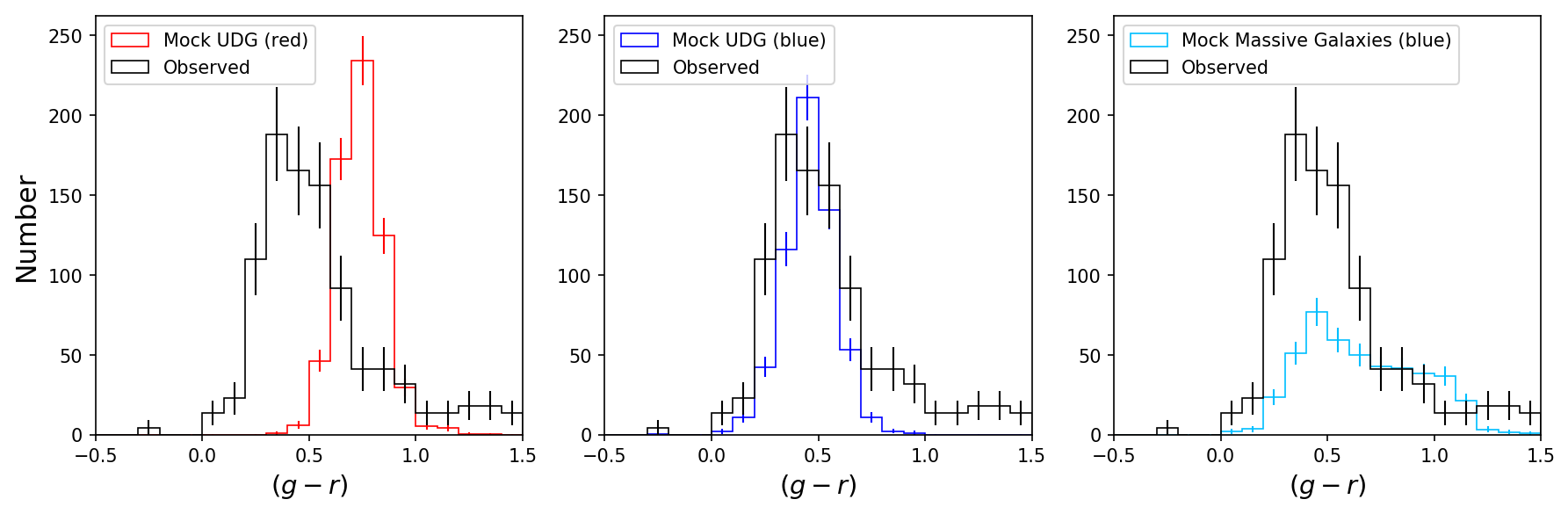}
	\centering
	\caption{Synthetic distributions of $(g-r)$ colour for our mock red UDG, blue UDG and massive blue galaxy catalogues, weighted by the probability of observation, compared to the actual observed histogram. The absolute numbers are normalised to an area of 180 square degrees. The error-bars show the Poisson uncertainties in each bin. Colours are in the observed reference frame. We note that we include the effect of measurement error in our mock colours. It is clear that the red UDG model is not consistent with the observations, being much more consistent with the blue model.}
	\label{figure:grplot}
\end{figure*}


\subsubsection{Canonical empirical distributions}
\label{section:Canonical}

\noindent While relatively little is known about the population of field galaxies with stellar masses lower than around $10^{9}\mathrm{M}_{\odot}$, much is known about objects at higher stellar masses. We can therefore produce mock observational catalogues of high stellar-mass galaxies and use them to estimate the contamination level in our actual observed catalogue. Such an estimate would be naturally conservative owing to the truncation at $10^{9}M_{\odot}$ that essentially excludes all dwarf galaxies including UDGs \citep[e.g.][]{Prole2019}. The ingredients of our model catalogues are:

\begin{itemize}
	\item The stellar mass function (SMF). We have used measurements from GAMA \citep{Baldry2012} and COSMOS/UltraVISTA \citep{Muzzin2013} to model the galaxy SMF of field galaxies, including its redshift dependence between $z$=0 and $z$=1. These measurements are additionally decomposed into red and blue galaxy populations and we have incorporated this into our mock catalogues.
	
	\item The stellar-mass to size relation. We used the measurements of \cite{vanderWel2014} to assign effective circularised radii to each of our random sample of stellar masses, as a function of redshift. Again, we make use of their early/late-type colour decomposition. We also include scatter in the sampling based on their measurements.
		
	\item We assign ($g-r$) colours to our sampled galaxies as a function of their stellar mass by using empirical data gathered by the  GAMA survey\footnote{Specifically, we use the public \texttt{StellarMasses v19} catalogue available from http://www.gama-survey.org/dr3/.} \citep{Taylor2011}. Specifically, we bin their $(g-r)$ measurements in stellar mass and assign intrinsic (i.e. observed at $z$=0) colours to our mock observations in each bin by randomly sampling from the corresponding GAMA ($g-r$) measurements.
	
	\item We calculate $k$-corrections in the same way as described in $\S$\ref{section:predict_UDG_cosmo}, this time assuming an onset of star formation 8.9 Gyr ago and [Z/H]=-0.38, with a uniform star formation rate. These values are based on the high surface brightness, late type sample of \cite{Ferre-Mateu2018}. The effects of modifying this are discussed in appendix \ref{appendix:uncertainty}.
	
	\item As before, we assume the redshift distribution of our catalogue to be smoothly distributed with the mass in the Universe, according to equation \ref{equation:Mtot}.
\end{itemize}

\indent Clearly such an exercise is approximate in nature, and is designed only to get a first-order estimate of the number of contaminants in our UDG sample. A discussion regarding the contribution to our observations from LSB galaxies smaller than the fiducial UDG limit of \re=1.5 kpc can be found in appendix \ref{appendix:extensions}. 

\indent Many of the mock massive galaxies are brighter than $m_{r}$=19.8. This means that it is possible to compare the redshift distribution of our mock catalogue with that of the GAMA spectroscopic survey. We show in appendix \ref{appendix:redshift} that our mock catalogues are consistent with that observed by GAMA.


\subsection{Measurement Errors}
\label{section:predict_error}

Before the catalogues can be directly compared with our observations, it is important to consider the effect of measurement errors on the predicted distributions of observed parameters. Of particular importance is the uncertainty in \re, which increases for larger and fainter galaxies. This is significant because there is typically a steep gradient in the distribution of \re, whereby there are far less large objects than small ones, both in terms of physical and angular size. Thus, including the measurement error in the mock catalogues causes an increase in the predicted number of galaxies observed with large angular sizes.

\indent The measurement uncertainty on the S\'ersic parameters is estimated directly from the synthetic source injections described in $\S$\ref{section:RE}. Measurements of the injected sources are used to estimate the recovery efficiency, defined in intrinsic observable parameter space. Consequently, the effects of measurement error (including any bias) as a function of intrinsic size and surface brightness are already contained in the recovery efficiency estimate. We can therefore account for the effect of the measurement uncertainty in our mock catalogues by using the recovery efficiency to assign probabilities of detection (see $\S$\ref{section:results_comp}). The limitation of this approach is that we cannot directly compare structural parameters in our mock catalogues with the observations.

\indent Also of importance is the measurement error in $(g-r)$ colour. Starting from our estimates of fluxes and their errors described in $\S$\ref{section:colours_obs}, we perform monte-carlo realisations of flux ratios in order to estimate the distribution of uncertainties in the magnitude. We fit a log-normal distribution to the result, and use it to randomly sample uncertainties in colour; we then ``jiggle'' (randomly perturb within error) the colours in the mock catalogues according to the result. The mean uncertainty in colour is $\sim$0.04 mag.


\section{Results}
\label{section:results}


\subsection{Observations vs. Model Predictions}
\label{section:results_comp}

We are now in a position to compare our mock catalogues with the observations. We note that for this analysis, all absolute numbers are normalised to an area of 180 square degrees. Each source in our mock catalogues is assigned a probability of recovery using the recovery efficiency discussed in $\S$\ref{section:RE}, which are used as weights in the analysis. We note that after using such weights, the number of UDGs we predict to observe converges (i.e. does not increase further) by $z$$\sim$0.2 (see figure \ref{figure:redshifts}). Similarly, the mock massive galaxy catalogue converges by $z$$\sim$0.5. This is mainly an effect of the lower limit angular size cut at \re$\geq3\arcsec$. We probe the accuracy of our modelling with reference to the redshift distribution of GAMA spectroscopic sources in appendix \ref{appendix:redshift}.

\indent We compare the $(g-r)$ histogram of our observed UDG candidate catalogue with each of our mock catalogues (red UDGs, blue UDGs, massive blue galaxies) in figure \ref{figure:grplot}; the results of which are fairly striking. Clearly either the assumption that all UDGs are on the red sequence as they are in clusters is not correct (as made clear by the significant offset between the peaks of the observed and predicted distributions), or UDGs in general do not form in the field with a mass-efficiency anywhere near what they do in clusters. However, since we already know that blue UDGs do exist in abundance in the field \citep[e.g.][]{Leisman2017, Jones2018}, it is clear that we can rule the latter hypothesis out completely. From this result we would expect isolated red UDGs, like the ones found by  \cite{Martinez-Delgado2016} and \cite{Roman2019}, to be relatively rare.

\indent The discrepancy is further compounded if one considers the estimates for the massive blue galaxy interlopers. We argue that since the massive blue galaxies are the dominant source of contamination in our UDG catalogue (see $\S$\ref{section:predict_interlopers} and appendix \ref{appendix:extensions}), we can obtain an observational sample that is representative of the UDG population by statistically subtracting the massive blue galaxy population from the observed catalogue of UDG candidates. The result is displayed in the left panel of figure \ref{figure:grplot_lit}, along with our mock UDG catalogues.

\indent It is clear that the mock blue UDG catalogue is in much better agreement with the observed colour distribution than the red UDG catalogue. However, the observations are $\sim$0.05 magnitudes bluer than what our empirical models predict. This means that the colours may be more consistent with the blue UDGs of \cite{Roman2017b} (see figure \ref{figure:model_colours}). This is not particularly surprising; late-type galaxies in clusters are typically redder than those in the field because of environmental quenching from e.g. ram-pressure stripping.

\indent By comparing our mock catalogues to the interloper-corrected observations, it is possible to estimate the total field density of UDGs, along with a corresponding mass formation efficiency. This is accomplished by comparing the predicted number of UDGs from our empirical model with the estimated number of observed UDGs. From the appearance of figure \ref{figure:grplot_lit} (left panel), it is clear that we have overestimated the number density (and therefore mass formation efficiency) of UDGs in our model.

\indent For the empirical model to predict the correct number of UDGs, we would require a mass-formation efficiency \UDGefficiency\ times that what it is in clusters, taking into account uncertainties discussed in appendix \ref{appendix:uncertainty}. This translates into a field density of \UDGdensity. This is an upper limit on the true field abundance of UDGs because the estimated number of observed UDGs is likely an overestimate; we have only considered massive blue galaxies as contaminant sources. We note that these estimates apply only to the range of physical parameters that we have probed here, i.e. sizes in the range 1.5$\leq$\rer[kpc]$\leq$7.0 and intrinsic (i.e. not cosmologically dimmed) surface brightnesses spanning 24.0$\leq$\uaer$\leq$26.5. If we were to consider even fainter sources, this number density would likely increase.


\subsection{Dependence on S\'ersic index}
\label{section:sersicn}

UDGs typically have S\'ersic indices $n$$<$1.5. The justification for our cut at $n$=2.5 is as follows:

\begin{enumerate}
	\item Our measurements are conflated by measurement error, which gets worse as a function of surface brightness. In figure \ref{figure:galfit} we show that we are nevertheless able to recover essentially all non-nucleated profiles with $n$$<$1.5 by imposing a cut at $n$$\sim$2.
	\item Some UDGs are nucleated \citep[e.g.][]{Venhola2017}. In this analysis, we have only fit single S\'ersic profiles. The presence of a nucleus can bias our recovery of $n$ by +30\%, so a higher cut than $n$=2 is justified to preserve completeness.
	\item We do not explicitly include the S\'ersic index distribution in our empirical model for background interlopers\footnote{although this could potentially be implemented in future studies}, which is statistically subtracted from the observational sample in the analysis. It is typical in the literature to take $n$=2.5 as the dividing line between ``early'' and ``late'' type samples \citep[e.g.][]{vanderWel2008, Vulcani2014, Vika2015}. Since we do not want to over subtract the interloper population, it is important to use a consistent cut for the sample selection. 
	\item In this analysis, we are striving to place upper-limits on quantities like the UDG field number density; this is motivated by the fact that our interloper subtraction is likely incomplete. Lowering our S\'ersic index selection cut would reduce the size of the observational sample and thus lower the inferred number density. In the interests of upper-limits, it is therefore prudent to keep a relatively high cut in $n$.
	\item Our results are closely compared with the work of \cite{vanderBurg2016}, who used an even higher cut at $n$=4.
\end{enumerate}

\indent In summary, while the cut at $n$=2.5 might be relatively high compared to the observed values of  $n$ for UDGs, we account for the resulting contamination in our observational sample using the empirical model. However, it is important to discuss the effects of varying the index cut on our results. Recall that with the cut at $n$=2.5, the upper-limit mass formation efficiency is estimated to be \UDGefficiency\ times that in clusters. If we instead take the cut at $n$=2.0, this drops to $\sim$50\% of the value for clusters. Here we have likely increased the purity of UDGs in our sample, but for the reasons given above it is difficult to quantify the effect on the completeness. If instead we drop to $n$=1.5, the formation efficiency estimate drops to $\sim$30\% of its value in clusters. However, it is likely that this value suffers from significant completeness effects and is an underestimate. 


\section{Discussion}
\label{section:discussion}

\subsection{Comparison with H\RNum{1}-bearing UDGs}

\begin{figure*}
	\includegraphics[width=\linewidth]{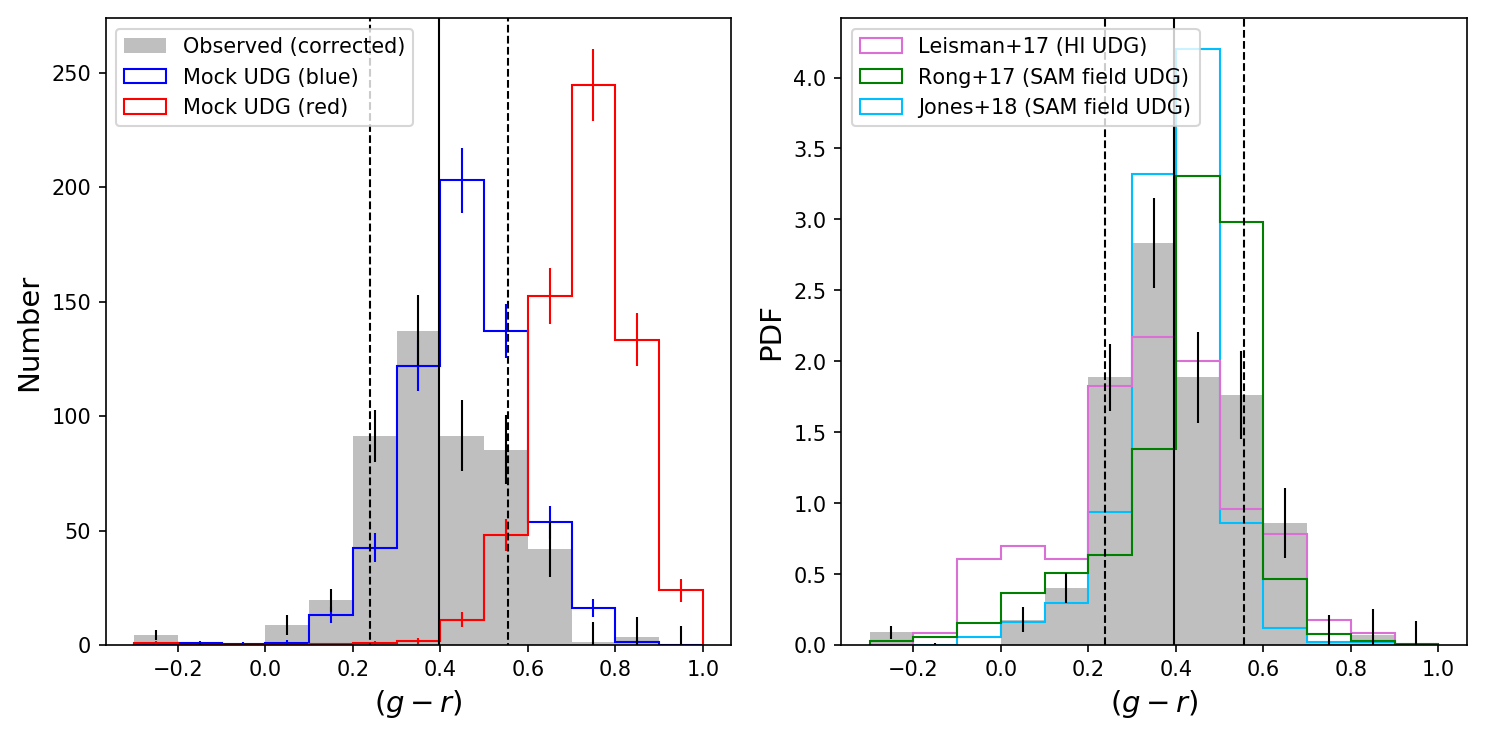}
	\centering
	\caption{The observed distribution of colour after subtracting the estimated contribution from massive blue galaxies (grey histogram). \textit{Left:} Comparison with the empirical red and blue UDG models from this study. We show the mean and 1$\sigma$ dispersion of the observations with the vertical lines. \textit{Right:} Normalised comparison with observations of H\RNum{1}-bearing field UDGs in the literature \protect\citep{Leisman2017}. and predictions from the semi-analytical models (SAM) of \protect\cite{Rong2017} and \protect\cite{Jones2018}. Poisson error-bars are shown. See text for discussion.}
	\label{figure:grplot_lit}
\end{figure*}

\indent In this section, we compare our observed, contaminant-corrected $(g-r)$ histogram to other measurements from the literature. We do not consider values of $(g-r)>1$ because they are almost certainly not part of the UDG population. One catalogue that we can directly compare with is that of \cite{Leisman2017}\footnote{We note that we use the HUD-B sample, which contains 115 sources and was selected using selection criteria consistent with that of \protect\cite{vanderBurg2016}.}, who measured the colours of isolated HI-bearing UDGs using SDSS data. While these measurements are conflated with measurement error because of the limited depth of SDSS, we can perform a qualitative comparison between the reported results (figure \ref{figure:grplot_lit}).

\indent In figure \ref{figure:grplot_lit}, we show how the colour distribution of our corrected observational sample of UDG candidates compares with that of \cite{Leisman2017}. The two PDFs are very similar, providing an indication that UDGs in the field are predominantly blue independently of the colour models we assumed in $\S$\ref{section:predict_UDG_colour}. There are some differences between the two distributions: we observe slightly more UDG candidates on the red side of the peak. There are several possible explanations: they could be sources that \cite{Leisman2017} were not sensitive to thanks to low H\RNum{1} content, or they are contaminant objects that we have not properly accounted for in our UDG sample such as massive early-type galaxies. Alternatively, since we are sensitive to UDGs out to $z\sim0.2$ comparing to their maximum distance of 120 Mpc ($z\sim0.03$), it may be that $k$-corrections play a role. The \cite{Leisman2017} catalogue also seems to have an excess of blue UDGs compared to what we observe. This could be explained either by the measurement error arising from the limited SDSS depth, or perhaps because blue UDGs are intrinsically brighter and we miss them in our selection (\cite{Leisman2017} use a slightly brighter bright-end selection cut). 

\indent \cite{Jones2018} used the \cite{Leisman2017} catalogue to estimate the field density of H\RNum{1}-bearing UDGs, obtaining a value of \UDGdensityHI. This is approximately one fifth of our upper-limit estimate of the overall UDG field density. However, comparing such field densities is difficult because the limited depth of the SDSS imaging used by \cite{Leisman2017} to identify UDGs creates significant measurement uncertainty, blurring their selection boundaries and leading to an uncertainty of $\sim$25\% in their sample size. However, using the fact that our estimate of the UDG density is an upper-limit, we can estimate that H\RNum{1}-bearing UDGs comprise at least one-fifth of the overall population.



\subsection{Comparison with Semi-Analytic Models}

\indent We also compare our results with the work of \cite{Jones2018}, who used the Santa Cruz semi-analytic model \citep[SAM,][]{Somerville2015} to generate a UDG sample in order to compare with the observations of \cite{Leisman2017}. Their results are also displayed in figure \ref{figure:grplot_lit}. We note that we jiggle their $(g-r)$ colours to match our measurement error for the comparison. The peak of their $(g-r)$ distribution is in reasonable agreement with our observations, yet it is slightly shifted towards the red and narrower. This may be because our observed catalogue is not entirely made from UDGs but also contains some contaminant sources, or perhaps because the SAM does not reproduce the correct amount of scatter for UDG colours. Alternatively, it may be a projection effect; our observed colours are in the observed frame and therefore are subject to $k$-corrections.

\indent A similar comparison can be made with the work of \cite{Rong2017}, who used the \cite{Guo2013} SAM to obtain a catalogue of simulated UDGs. We again jiggle their colours using our measurement error for the comparison. The colour distribution of their field UDGs is shifted towards the red compared to our observations, as also noted by \cite{Jones2018}. If we were to include additional reddening of their colours because of $k$-corrections (i.e. to make a fair comparison with our observations), then this discrepancy would be exaggerated.

\indent We calculate the total field density of UDGs in the Santa-Cruz SAM by integrating the stellar-mass function for UDGs \citep[figure 4 of][]{Jones2018}. We obtain a value of \UDGdensitySAM, approximately twice the upper-limit estimate from our measurements.


\subsection{Impact of Nearby Galaxy Groups}

While our observed catalogue of UDG candidates is dominated by field sources, it is important to consider the effects of nearby galaxy groups on our result. After all, if such sources are predominantly quiescent and exist in similar number to our field sample, we should expect to find a population of red UDG candidates.

\indent Fortunately, our KiDS/HSC-SSP footprint overlaps with the GAMA spectroscopic survey \citep{Driver2011} and thus the GAMA group catalogue \citep{Robotham2011}. We can therefore make an estimate of the contribution of group/cluster UDGs to our field sample using a similar method to \cite{vanderBurg2017}. Working in our favour is the fact that we have applied a relatively high cut in angular size (\re$\geq$3$\arcsec$) and there are no massive groups that are close enough to dominate our selection.

\indent We select all GAMA groups from the group catalogue that satisfy \texttt{Nfof}$\geq$5 and 0.01$<$\texttt{Zfof}$<$0.2 (where \texttt{Nfof} and  \texttt{Zfof} are respectively the number of friends-of-friends sources and an estimate of the group redshift). For each group, we count the number of sources that are compatible with being UDGs at that redshift, using our selection criteria and a physical radius range of 1.5$\leq$\re\ [kpc]$\leq$7.0. We also subtract a statistical background correction based on the total number of sources across the entire footprint that meet the same criteria. We estimate that up to 8\% of our UDG candidate catalogue is associated with relatively massive groups (i.e. the ones that have at least five friends-of-friends members), with the uncertainty coming from the background count estimate. The colour histogram of these sources is statistically indistinguishable from that of the whole catalogue, and we conclude that their inclusion does not significantly impact our results.

\indent Using the empirical scaling relation between the group mass and total $r$-band luminosity from \cite{Viola2015}, we estimate that $\sim6\%$ of the available mass out to $z=0.2$ is taken up by the groups we consider here. This is very similar to the fraction of observed UDG candidates associated with groups. Taken with the fact that we expect all observed UDGs to be at $z<0.2$, this provides an independent indication that the mass formation efficiency of UDGs in the field is comparable to that in groups and clusters. It also shows that the presence of the massive groups does not severely impact our result.

\label{section:groups}




\section{Conclusions}
\label{section:conclusion}

In this paper we have used deep, wide-area optical imaging from the KiDS survey to detect sources with low surface brightness (\SBsel) and large angular sizes (\resel). Following the detection and measurement of these sources with \MTObjects\ and \GALFIT, we measured colours using the HSC-SSP survey data. Our catalogue of UDG candidates consists of \Nsources\ sources over $\sim$\UnmaskedArea\ square degrees. Compared to UDGs in groups and clusters, our sample consists of sources that appear to have much more irregular morphologies and show hints of active star formation.

\indent These observations were compared to mock observations of UDGs created by sampling empirical distributions of UDG properties from the literature. Our key assumptions were intrinsic size, surface brightness and colour distributions for the UDGs. All the assumptions we made are justified based on the current understanding of UDGs.

\indent By comparing our mock catalogues with the observations, we have shown that it is very unlikely for a significant population of UDGs that are as red in colour as they are in clusters to exist in the field. It is much more likely that almost all UDGs in the field are instead much bluer, with colours similar to late-type dwarf galaxies in clusters. An immediate conclusion based on the predominantly blue colours is that secular evolutionary processes are not producing large numbers of cluster-like quenched (red) UDGs. 


\indent This finding means that isolated red UDGs, like the ones found by \cite{Martinez-Delgado2016} and \cite{Roman2019}, should be quite rare. At first glance this contrasts with the work of \citep{Pranjal2019}, who find a population of UDGs with low specific star formation rates (compared to the star forming main sequence) in the field. Taking the appearance of our detections in figure \ref{figure:examples} into account, it is likely that UDGs in the field are forming stars only in a few localised sites; this locality may result in relatively low specific star formation rates that nevertheless make their integrated colours bluer. As noted by \citep{Pranjal2019} and \citep{Zaritsky2018}, their sample of field UDGs is still systematically bluer than that observed for cluster UDGs. Assuming that UDGs across different environments share similar metallicities, this is good evidence that star formation in field UDGs can be quite tentative and easily quenched in cluster environments. 

\indent We also created mock observations of massive blue galaxies, thought to be the primary source of contamination in our UDG candidate sample, using canonical empirical relations. We statistically subtracted these from our observations to acquire a contaminant-corrected catalogue of UDGs. The normalised distribution of $(g-r)$ colour is very similar to that estimated for H\RNum{1} bearing field UDGs measured by \cite{Leisman2017}. The observed distribution is also similar to that predicted for UDGs in SAMs \citep{Rong2017, Jones2018}, but slightly bluer. While our colour distribution appears to have greater dispersion, this is likely due to systematic shortcomings in comparing simulations with observations.

\indent Using our mock catalogues as a reference, we estimate an upper limit on the field density of UDGs as \UDGdensity, equivalent to a mass formation efficiency \UDGefficiency\ times that in clusters. Perhaps surprisingly, this density actually implies that UDGs form with a mass efficiency in the field that is quite close to what they do in cluster environments. The field density applies for UDGs with physical sizes 1.5$\leq$\rer[kpc]$\leq$7.0 and intrinsic (i.e. not cosmologically dimmed) surface brightnesses 24.0$\leq$\uaer$\leq$26.5. This number density also suggests that current SAMs over-predict the number of UDGs by at least a factor of two. However, we note that if UDGs exist in abundance at lower surface brightnesses than what we have probed here, the total number density of large LSB objects could be much higher. Based on the field density measured by \cite{Jones2018}, H\RNum{1}-bearing UDGs comprise at least one fifth of the overall UDG population in the field. This is consistent with what is predicted from the Santa-Cruz SAM.

\indent We note that the analysis we have performed in this work has been approximate in nature because of the absence of any distance measurements. Acquiring large samples of spectroscopic redshifts for LSB galaxies in the field is not currently feasible. In the near-term the second data-release of the HSC-SSP will provide an opportunity to follow-up the present work thanks to its expanded footprint; this analysis can easily be expanded to larger areas. In the longer term, deep all-sky imaging (perhaps combined with photometric redshifts) from LSST may provide the ultimate data set for providing statistical constraints on LSB galaxies in the field.






\section*{Acknowledgements} 

\noindent We are thankful to Caroline Haigh and the \MTObjects\ team for providing their software.\newline

\noindent We are also grateful to Javier Rom\'an, Yu Rong and Michael G. Jones (and collaborators) for kindly sharing their data.\newline

\noindent  We are additionally thankful to Alister Graham, Ivan Baldry and Benne Holwerda for providing insightful comments and suggestions.\newline

\noindent We would also like to acknowledge the contribution that the reviewer, Javier Rom\'an, has made to the quality of the paper.\newline

\noindent Based on data products from observations made with ESO Telescopes at the La Silla Paranal Observatory under programme IDs 177.A-3016, 177.A-3017 and 177.A-3018, and on data products produced by Target/OmegaCEN, INAF-OACN, INAF-OAPD and the KiDS production team, on behalf of the KiDS consortium. OmegaCEN and the KiDS production team acknowledge support by NOVA and NWO-M grants. Members of INAF-OAPD and INAF-OACN also acknowledge the support from the Department of Physics \& Astronomy of the University of Padova, and of the Department of Physics of Univ. Federico II (Naples).\newline

\noindent GAMA is a joint European-Australasian project based around a spectroscopic campaign using the Anglo-Australian Telescope. The GAMA input catalogue is based on data taken from the Sloan Digital Sky Survey and the UKIRT Infrared Deep Sky Survey. Complementary imaging of the GAMA regions is being obtained by a number of independent survey programmes including GALEX MIS, VST KiDS, VISTA VIKING, WISE, Herschel-ATLAS, GMRT and ASKAP providing UV to radio coverage. GAMA is funded by the STFC (UK), the ARC (Australia), the AAO, and the participating institutions. The GAMA website is http://www.gama-survey.org/.\newline

\noindent This research made use of Astropy,\footnote{http://www.astropy.org} a community-developed core Python package for Astronomy \citep{astropy2013, astropy2018}.\newline

\noindent This research has made use of the VizieR catalogue access tool, CDS, Strasbourg, France (DOI : 10.26093/cds/vizier). The original description of the VizieR service was published in A\&AS 143, 23.\newline

\noindent This research has made use of NASA's Astrophysics Data System.


\bibliographystyle{mnras}
\bibliography{library.bib}

\appendix


\section{Examples}

\begin{figure*}
	\includegraphics[width=\linewidth]{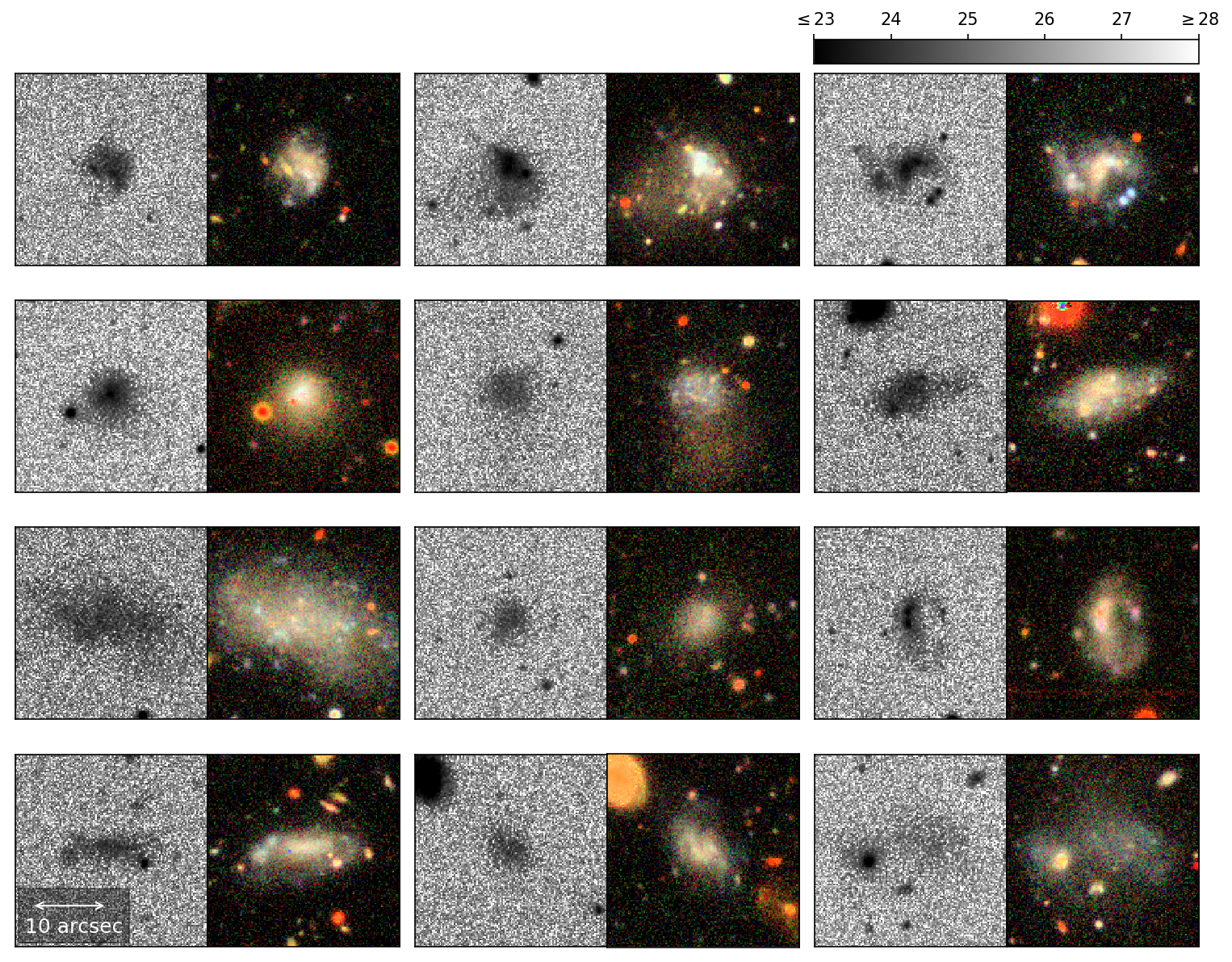}
	\centering
	\caption{Examples of large LSB sources in our UDG candidate sample. Each source in shown in KiDS $r$-band (left panels) vs. a colour image made from the $g$, $r$ \& $i$ HSC-SSP bands according to \protect\cite{Lupton2004} (right panels). The cut-out size is 25$\arcsec$, much less than the 80$\arcsec$ regions that we use to fit the sources. The colour bar shows the surface brightness in units of magnitudes per square-arcsecond for the KiDS data. In comparison to the regular morphologies of UDGs in groups and clusters, many of our sources are rather amorphous, with possible signs of discrete sites of active star formation. }
	\label{figure:examples}
\end{figure*}

\label{appendix:examples}


\section{Model uncertainties}
\label{appendix:uncertainty}

\begin{figure}
	\includegraphics[width=\linewidth]{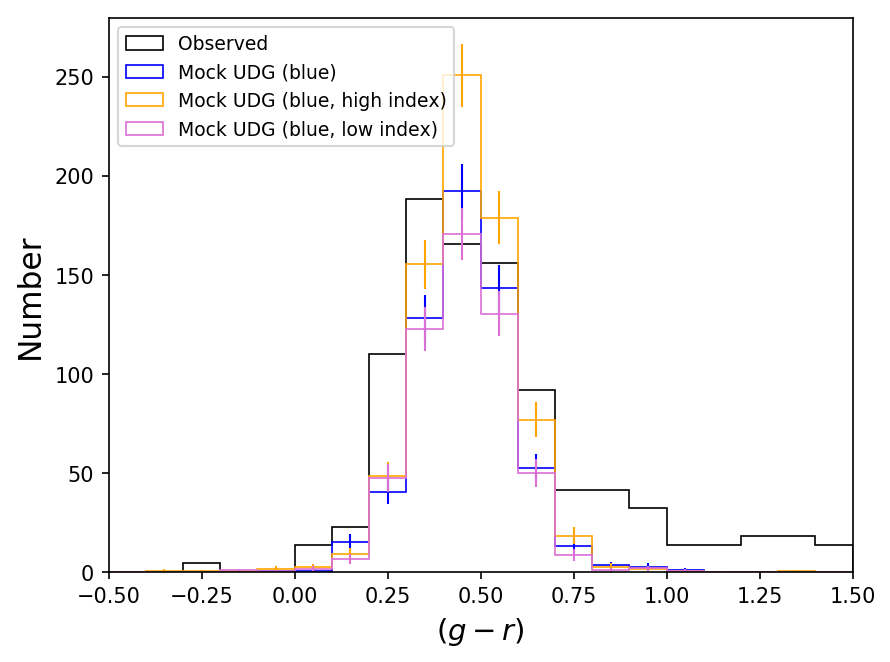}
	\centering
	\caption{Impact of changing model parameters within their errors. The observed colour histogram is shown in back. The blue UDG model is shown in dark blue. The mock catalogue with the size power law index lowered/raised by $1\sigma$ is shown in purple/orange. All error-bars are Poisson uncertainties.}
	\label{figure:uncertainties}
\end{figure}

While our analysis here is first order in nature, it is still important to quantify how uncertainties in the model ingredients may impact our result. In particular, we have not discussed how the uncertainties in the assumed UDG size distribution propagates. \cite{vanderBurg2017} have measured a power-law index of -2.71$\pm$0.33 for the distribution of circularised radii (equation \ref{equation:remodel}). The result of varying the slope by 1$\sigma$ are shown in figure \ref{figure:uncertainties}. It is clear that lowering the index (more small UDGs) lowers our estimate of the number of UDGs we expect to observe by around 15\%. Conversely, increasing the index (more large sources) causes the predicted number to increase by around 25\%.

\indent An additional source of uncertainty is that which arises from our estimate of the mass formation efficiency. From the empirical relation of \cite{vanderBurg2017}, we estimate a $\sim$20\% error. This uncertainty also propagates to our estimate of the field density of UDGs.

\begin{figure}
	\includegraphics[width=\linewidth]{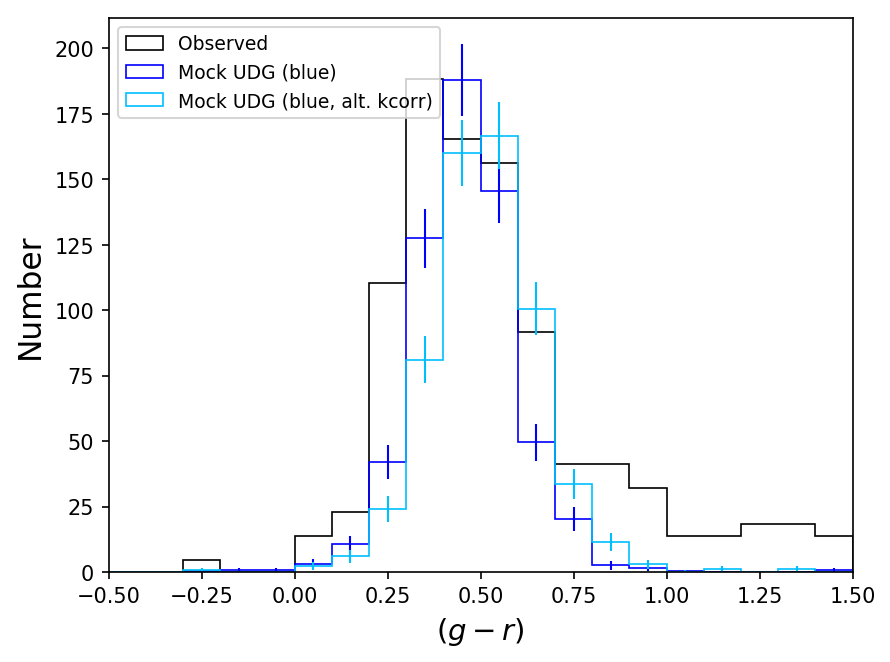}
	\centering
	\caption{Comparison of the model colour distributions for blue UDGs assuming a star-forming stellar population for the $k$-corrections (dark blue), and a quiescent one (light blue). Overall, we find the difference to be negligible for our analysis.}
	\label{figure:altk}
\end{figure}

\indent A separate issue is how the assumed stellar population (i.e. that which defines the $k$-corrections) affects our analysis. For the UDGs, we have explored a red and blue colour model, using quiescent and star-forming populations for the $k$-corrections respectively. One of the uncertainties for the star-forming population model is the star formation history to assume; for this analysis we have assumed a uniform star formation rate. As a means to test whether this assumption impacts our result, we can also model the blue UDG population using the quiescent model for $k$-corrections. The results of this are shown in figure \ref{figure:altk}. From this figure, it appears that the change is small, with a slight shift towards redder colours. The impact on our analysis is negligible; this is not surprising as most of our observed UDGs are expected to be at low redshift where $k$-corrections are small.

\indent We have repeated a similar process for our late type interloper model, replacing the stellar population model with that used for star-forming UDGs. We also find that this makes no significant difference to our results.

\section{Extending the model}

One shortcoming of our analysis is that we do not account for galaxies smaller than the fiducial UDG size limit of 1.5 kpc, a fairly arbitrary cut-off. One way to probe how the inclusion of such galaxies may alter the results presented in figure \ref{figure:grplot} is to extrapolate the empirical size distribution that we use in our UDG model \citep{vanderBurg2017} to lower size limits. This exercise is approximate in nature because it is not clear whether an extrapolation of this relation is valid for smaller galaxies. There are two competing effects: while smaller galaxies are more numerous because of the steep power law (equation \ref{equation:remodel}), their smaller size means that they are much less likely to be observed given our selection criteria and the corresponding recovery efficiency.

\begin{figure}
	\includegraphics[width=\linewidth]{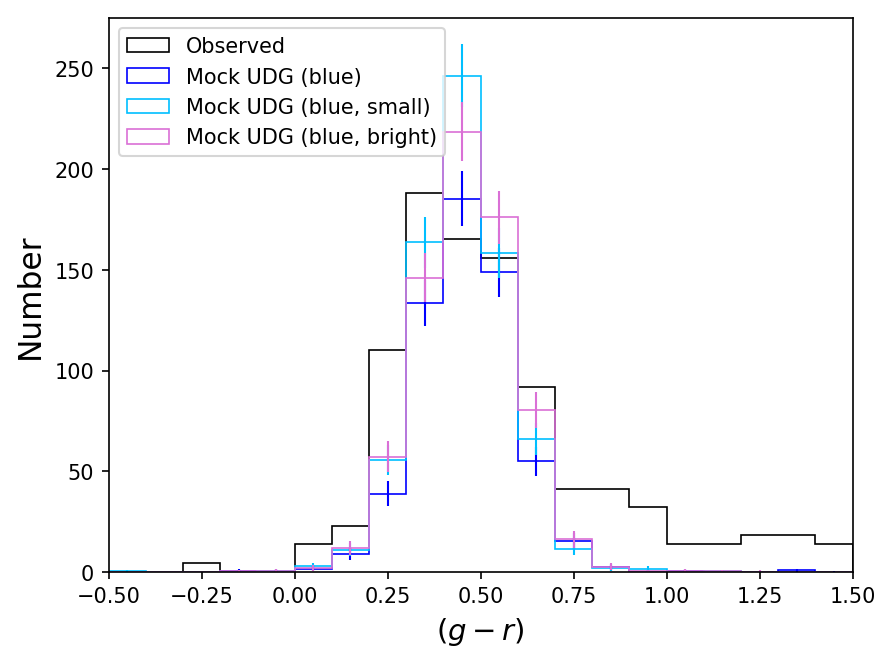}
	\centering
	\caption{Results of extending our empirical model. The observed colour histogram is shown in back. The blue UDG model is shown in dark blue. The mock catalogue with the reduced lower limit of \re$\geq$0.5 kpc is shown in light blue. The mock catalogue that includes bright galaxies (\uae$\geq$22) is shown in purple. All error-bars are Poisson uncertainties.}
	\label{figure:extensions}
\end{figure}

\indent We display the result of reducing the lower physical size limit from 1.5 kpc to 0.5 kpc in figure \ref{figure:extensions}. The total number of sources we generate are increased proportionally to equation \ref{equation:remodel}. Despite the increase in the number of sources (by a factor of $\sim$20), there is only a minor difference between the number of sources we would expect to observe.

\indent As a further extension to the model, we can consider what happens when we decrease the lower limit in surface brightness at \uae=24 to allow brighter sources into the selection. Since \cite{Danieli2018} have shown that the distribution of intrinsic size of large red galaxies is approximately uniform with absolute magnitude (and therefore surface brightness), this extension can be interpreted as including large red galaxies with S\'ersic indices meeting our selection criteria. As an example, we show in the figure the effect of using a bright-end surface brightness cut of \uae=22, increasing the number of sources by 80\% according to equation \ref{equation:sbmodel}. As in the previous test, the difference with the result in figure \ref{figure:grplot} is fairly insignificant. We are left to conclude that our observational sample is indeed likely made up of large low surface brightness galaxies.


\label{appendix:extensions}


\section{Comparison with GAMA redshifts}


We can compare our measurements and mock catalogues against measurements from the GAMA spectroscopic survey in order to test how well our mock catalogues represent reality. For this test, we use our best model: the combination of blue UDGs with massive blue galaxy interlopers. We assume that UDGs form with a mass efficiency as calculated in $\S$\ref{section:results_comp}.

\begin{figure}
	\includegraphics[width=\linewidth]{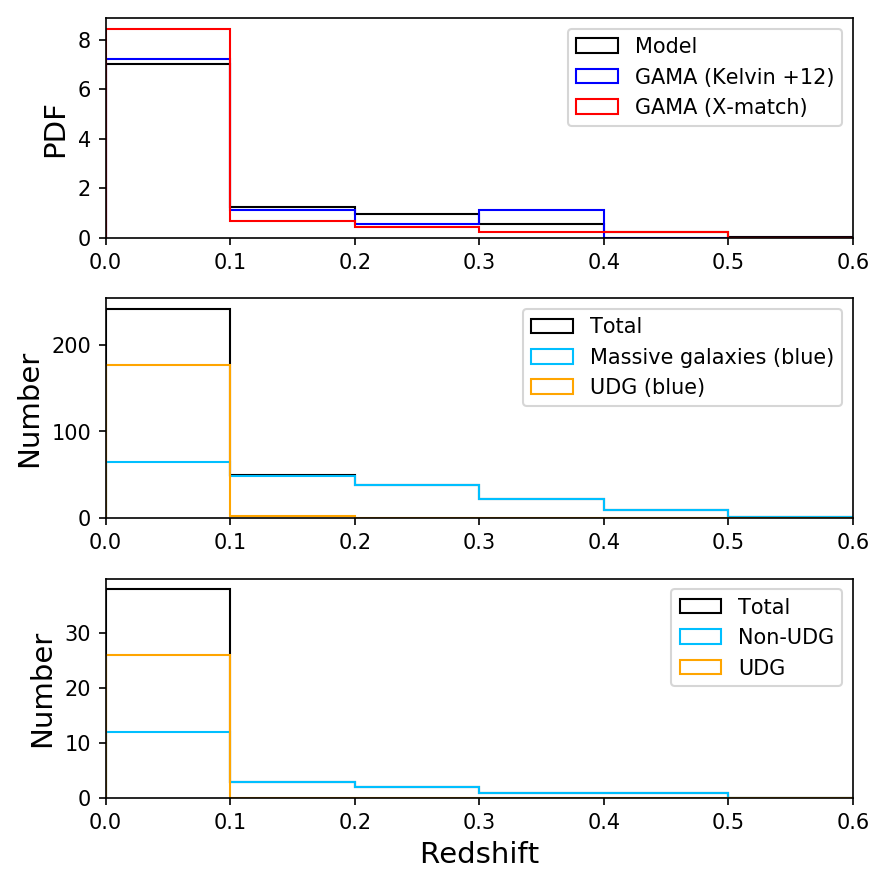}
	\centering
	\caption{\textit{Top:} Comparison of the overall redshift distribution from our mock catalogue (black histogram) with that of \protect\cite{Kelvin2012} (blue histogram) and a crossmatch between our observed UDG candidate catalogue and the GAMA spectroscopic survey (red histogram). \textit{Middle:} Decomposition of our mock catalogue into UDGs (orange) and interlopers (blue) as a function of redshift. \textit{Bottom:} The same as the middle panel, but for our crossmatch with GAMA.}
	\label{figure:redshifts}
\end{figure}

Using the public data release 3 data obtained from the GAMA website\footnote{http://www.gama-survey.org/}, we crossmatched the \texttt{SpecObj v27} catalogue (containing spectroscopic redshifts) with the \texttt{SersicCatSDSS v09} table \cite[containing S\'ersic profile fits to GAMA targets in SDSS data from][]{Kelvin2012}. We imposed our selection criteria on the S\'ersic parameters and additionally required \texttt{SURVEY\_CLASS}$\geq$4 in order to select legitimate sources with $m_{r}$$<$19.8, leaving us with 209 GAMA sources. We also applied the $m_{r}$$<$19.8 criterion to our mock catalogue, retrieving 379 sources. The results of the comparison are shown in the top panel of figure \ref{figure:redshifts}.

\indent Despite the surface brightness limits of GAMA \citep[e.g.][]{Wright2017}, we find that 45 of our UDG candidates (over the full unmasked KiDS area) have matches in the GAMA \texttt{SpecObj} catalogue within 3$\arcsec$. This allows us to make the same comparison as above, using our measurements in place of those of \cite{Kelvin2012}; the result is also shown in the top panel of figure \ref{figure:redshifts}. Clearly the overall distribution of our mock catalogues is in good agreement with the observations.

\indent However, we can go one step further; We can now estimate the intrinsic parameters of the matching sources because we know their redshifts and apparent structural parameters. In doing so, we can test whether the individual distributions for UDGs and interlopers are approximately correct. For this test, we define a UDG as having \re$\geq$1.5 kpc and $M_{*}$ $\leq$ $10^{9}\mathrm{M}_{\odot}$. The stellar mass was estimated assuming our blue UDG stellar population model from \ref{section:predict_UDG_cosmo} together with the \GALFIT\ $m_{r}$ measurements. We can then decompose the catalogues into UDG and non-UDG populations.

\indent The decomposition of the mock catalogue is shown in the middle panel of figure \ref{figure:redshifts}. This compares with the decomposed observed catalogue in the lower panel. Clearly the distributions are similar; at low redshifts UDGs dominate our sample, while at higher redshifts, massive interlopers dominate. 



\label{appendix:redshift}


\end{document}